\documentclass[fp,twocolumn]{jpsj3}
\usepackage{txfonts}
\usepackage{bm}
\usepackage{color}

\title{Sommerfeld--Bethe Analysis of $ZT$ in Inhomogeneous Thermoelectrics}

\author{
    Manaho Matsubara$^1$\thanks{mmatsubara@rs.tus.ac.jp}, Takahiro Yamamoto$^{1,2}$ and Hidetoshi Fukuyama$^{2}$}
    \inst{$^1$Department of Physics, Faculty of Science, Tokyo University of Science, Shinjuku, Tokyo 162-8601, Japan \\
    $^2$ Research Institute for Science and Technology, Tokyo University of Science, Shinjuku, Tokyo 162-8601, Japan} 

\abst{
  The development of good thermoelectric materials that exhibit high $ZT$ $\left(=\frac{PF}{\kappa} T\right)$ requires maximizing the power factor, $PF$, which is mainly governed by electrons, and minimizing the thermal conductivity, $\kappa$, which is associated with not only electrons but also phonons.
  In the present work, we focus on GeTe and Mg$_3$Sb$_2$ as high-$ZT$ materials with inhomogeneous structures and analyze both their electrical conductivity, $L_{11}$, and Seebeck coefficient, $S$, with help of Sommerfeld--Bethe formula, resulting in understanding the temperature dependence of $PF$ and the identification of electronic contribution to thermal conductivity, $\kappa_{\rm el}$.
  Comparing the obtained $\kappa_{\rm el}$ and experimentally measured $\kappa$ values, the temperature dependence of phononic contribution to thermal conductivity, $\kappa_{\rm ph}=\kappa-\kappa_{\rm el}$, is inferred and analyzed on the basis of the Holland formula.
  A comparison of the GeTe and Mg$_3$Sb$_2$ with different morphologies (i.e., GeTe exhibits a semi-ordered zigzag nanostructure similar to a disrupted herringbone structure,
  whereas Mg$_3$Sb$_2$ exhibits a rather uniform amorphous structure) reveals that
  the effect of grain size and defects on the temperature dependence of $\kappa_{\rm ph}$ is large in the former but small in the latter.
  Hence, we concluded that not only the size of the grains but also their shape strongly influences $\kappa_{\rm ph}$ and, consequently, $ZT$.}

\begin{document}
\maketitle

\section{Introduction}
In order to develop effective thermoelectric materials suitable for use under versatile conditions one needs solid scientific understanding based on electronic states as seen in the history of semiconductor technology. 
For such efforts, a detailed theoretical analysis of experiments demonstrating particular features with high values of figures of merit $PF$ (power factor) and $ZT$ (dimensionless thermoelectric figure of merit) will be useful because the results will lead to the identification of important factors for the further development of good thermoelectrics.
Here, the basic theoretical framework for this analysis is the Kubo--Luttinger theory for thermoelectricity~\cite{rf:kubo,rf:luttinger}, which gives the electrical conductivity, $L_{11}$, thermoelectrical conductivity, $L_{12}$, electrothermal conductivity, $L_{21}$, and thermal conductivity (for zero electric field), $L_{22}$, by correlation functions between the current density, $J_{\rm el}$, and the thermal current density, $J_{\rm th}$:
\begin{align}
    L_{11}&: \int_{0}^\beta \!\!\!d\tau\left\langle T_\tau \{J_{\rm el}(\tau)J_{\rm el}(0)\}\right\rangle e^{i\omega_\lambda\tau},\\
    L_{12}=L_{21}&: \int_{0}^\beta \!\!\!d\tau\left\langle T_\tau \{J_{\rm el}(\tau)J_{\rm th}(0)\}\right\rangle e^{i\omega_\lambda\tau},\\
    L_{22}&: \int_{0}^\beta \!\!\!d\tau\left\langle T_\tau \{J_{\rm th}(\tau)J_{\rm th}(0)\}\right\rangle e^{i\omega_\lambda\tau},
\end{align}
where $\beta$ is the inverse temperature, $T_\tau$ is the imaginary-time-ordering operator, $\langle\cdots\rangle$ is the thermal average, and $\omega_\lambda$ is the boson Matsubara frequency.
In terms of these $L_{ij}$, $PF$ and $ZT$ are given by
\begin{eqnarray}
    PF=\frac{1}{T^2}\frac{L_{12}^2}{L_{11}},
    \label{eq:PF}
\end{eqnarray}
and
\begin{eqnarray}
    ZT= \frac{PF}{\kappa}T,
    \label{eq:PF}
\end{eqnarray}
respectively, where $T$ is temperature and $\kappa=\left(L_{22}-\frac{L_{12}L_{21}}{L_{11}}\right)\frac{1}{T}$ is the thermal conductivity (for zero electric current) consisting of two different sources: electrons $\kappa_{\rm el}$ and phonons $\kappa_{\rm ph}$. (Possible contributions to $\kappa$ by drag effects between electrons and phonons are included in $\kappa_{\rm ph}$, although they are expected to be small in general cases because of the necessity of extra coupling between them.)
Each $L_{\rm ij}$ can be assessed using thermal Green's functions.
Regarding the $PF$, which is mainly governed by electrons, we have recently analyzed the remarkable experimental finding of a large $PF$ by Shimizu et al.~\cite{rf:fese_shimizu} in high-quality FeSe thin films. This analysis led us to propose an interplay between two two-dimensional bands nearby located in energy: one with a Fermi level leading to a large $L_{11}$ and the other above the Fermi level contributing to a large value of the Seebeck coefficient, $S=L_{12}/(TL_{11})$, as in semiconductors~\cite{rf:fese_matsubara}. 
Incidentally, Uematsu et al. pointed out a similar possibility through studies on the interface of GaAs/AlGaAs~\cite{rf:gaas_uematsu}.

In the present paper, we address $ZT$, where the interplay between electrons and phonons is critical, unlike the case for $PF$. 
The examples we study in the present paper are GeTe-~\cite{rf:gete_mori} and Mg$_3$Sb$_2$-based compounds~\cite{rf:mgsb_imasato}. Both of these materials exhibit a high $ZT$, reflecting a low $\kappa$, and they share similar features of highly inhomogeneous structures.
However, they show different relations between such structures and $\kappa$ (i.e., $\kappa$ is sensitive to changes in the defect structure in GeTe but insensitive to changes in the grain size in Mg$_3$Sb$_2$).

We first analyze $PF$ which results in possible values of the spectral conductivity, $\alpha$, in $L_{11}$ and $L_{12}$ with the help of Sommerfeld--Bethe formula~\cite{rf:sommerfeld}.
The thus-determined $\alpha$ value leads naturally to identification of the electronic contribution to the thermal conductivity, $\kappa_{\rm el}$. By comparison with experimentally observed $\kappa$ values, the phononic thermal conductivity, $\kappa_{\rm ph}$, can be identified by $\kappa_{\rm ph}$ = $\kappa$ - $\kappa_{\rm el}$.
The magnitude and the $T$ dependence of $\kappa_{\rm ph}$ yield information about the long-wavelength acoustic (LA) modes.
The results indicate that the damping of LA modes is sensitive to the morphology (e.g., the shapes and spatial extent of grain boundaries). This identification of properties of phonons will lead to further explorations of choices of materials and processes governing morphology.

Cases of GeTe- and Mg$_3$Sb$_2$-based compounds are studied separately in Sections~\ref{sec:gete} and \ref{sec:mgsb}, respectively, and these two cases are compared in Section~\ref{sec:comparison}. The results are summarized in Section~\ref{sec:summary}.

\section{Case of GeTe~\label{sec:gete}}

\subsection{Electronic States and Model of Spectral Conductivity}
{Pristine GeTe undergoes a phase transition from a rhombohedral to a cubic structure at $\sim$700~K, accompanied by essential changes in the band structure~\cite{rf:gete_transition1,rf:gete_transition2}, making systematic analysis not transparent. 
Hence, we focus on the rhombohedral structure in the present study,}
where the carriers (holes) are located in six valleys with an average effective mass $m^* = 0.25m_0$ ($m_0$ is the free electron mass)~\cite{rf:gete_mass2}.
$L_{11}$, $L_{12}$, and $L_{22}^{\rm el}$ (i.e., the electronic contribution to $L_{22}$) can be expressed by the Sommerfeld--Bethe formula as follows~\cite{rf:sommerfeld,rf:mahan,rf:cnt_bipolar,rf:sb_ogata}:
\begin{align}
    L_{11}&=\int_{-\infty}^\infty \!\!\!d\varepsilon\left(-\frac{\partial f(\varepsilon-\mu)}{\partial \varepsilon}\right)\alpha(\varepsilon),
    \label{eq:L11}\\
    L_{12}&=-\frac{1}{e}\int_{-\infty}^\infty \!\!\!d\varepsilon\left(-\frac{\partial f(\varepsilon-\mu)}{\partial \varepsilon}\right)(\varepsilon-\mu)\alpha(\varepsilon),
    \label{eq:L12}\\
    L_{22}^{\rm el}&=\frac{1}{e^2}\int_{-\infty}^\infty \!\!\!d\varepsilon\left(-\frac{\partial f(\varepsilon-\mu)}{\partial \varepsilon}\right)(\varepsilon-\mu)^2\alpha(\varepsilon),
    \label{eq:L22}
\end{align}
where $e$ is the elementary charge, and $f(\varepsilon-\mu) = 1/(\exp(\beta(\varepsilon-\mu)) + 1)$ is the Fermi--Dirac distribution function with the chemical potential, $\mu$, and where the spectral conductivity, $\alpha(\varepsilon)$, is given as~\cite{rf:cnt_bipolar,rf:sb_ogata} 
\begin{align}
    \alpha(\varepsilon)&=g_sg_v\frac{\hbar}{2\pi V}\sum_{\bm k}
    {\rm Re Tr}\left[j_x({\bm k})G^{\rm A}({\bm k},\varepsilon)j_x({\bm k})G^{\rm R}({\bm k},\varepsilon)\right.\notag\\
    &\left.\ \ \ -j_x({\bm k})G^{\rm R}({\bm k},\varepsilon)j_x({\bm k})G^{\rm R}({\bm k},\varepsilon)
    \right].
\end{align}
Here, $g_s(=2)$ and $g_v(=6)$ are the spin and valley degrees of freedom, respectively, $V$ is the system volume, $j_x({\bm k})$ is the current density in the $x$-direction along the $T$ gradient of the material, and $G^{\rm R/A}({\bm k},\varepsilon)$ is the retarded/advanced Green's functions.
Within the effective mass approximation and the constant-$\tau$ approximation, $G^{\rm R/A}({\bm k},\varepsilon)$ is represented as
\begin{align}
  G^{\rm R/A}({\bm k},\varepsilon)=\frac{1}{\varepsilon-\frac{\hbar^2k^2}{2m^*}\pm i\Gamma},
\label{eq:g}
\end{align}
where $\Gamma$ is the scattering rate, and thus $\alpha(\varepsilon)$ is rewritten as
\begin{align}
    \alpha(\varepsilon)=g_sg_v\frac{e^2\sqrt{m^*}}{12\pi^2\hbar^2}\frac{\left(\sqrt{\varepsilon^2+\Gamma^2}+|\varepsilon|\right)^{3/2}}{\Gamma},
    \label{eq:alpha}
\end{align}
where the energy origin is set at the band edge in a clean system.
The $\Gamma$ is assumed to be given by the Matthiessen's rule, $\Gamma(T) = \Gamma_{\rm imp} + \Gamma_{\rm el-ph}(T)$, where $\Gamma_{\rm imp}$ is due to impurity scattering and $\Gamma_{\rm el-ph}(T)$ is due to phonon scattering at high $T$, $\Gamma_{\rm el-ph}(T) = \gamma_{\rm el-ph}T$.

\subsection{Parameter Fitting for $L_{11}$, $S$, $PF$, and Electronic Contributions to $ZT$}
\begin{figure}[t]
    \begin{center}
        \includegraphics[keepaspectratio=true,width=76mm]{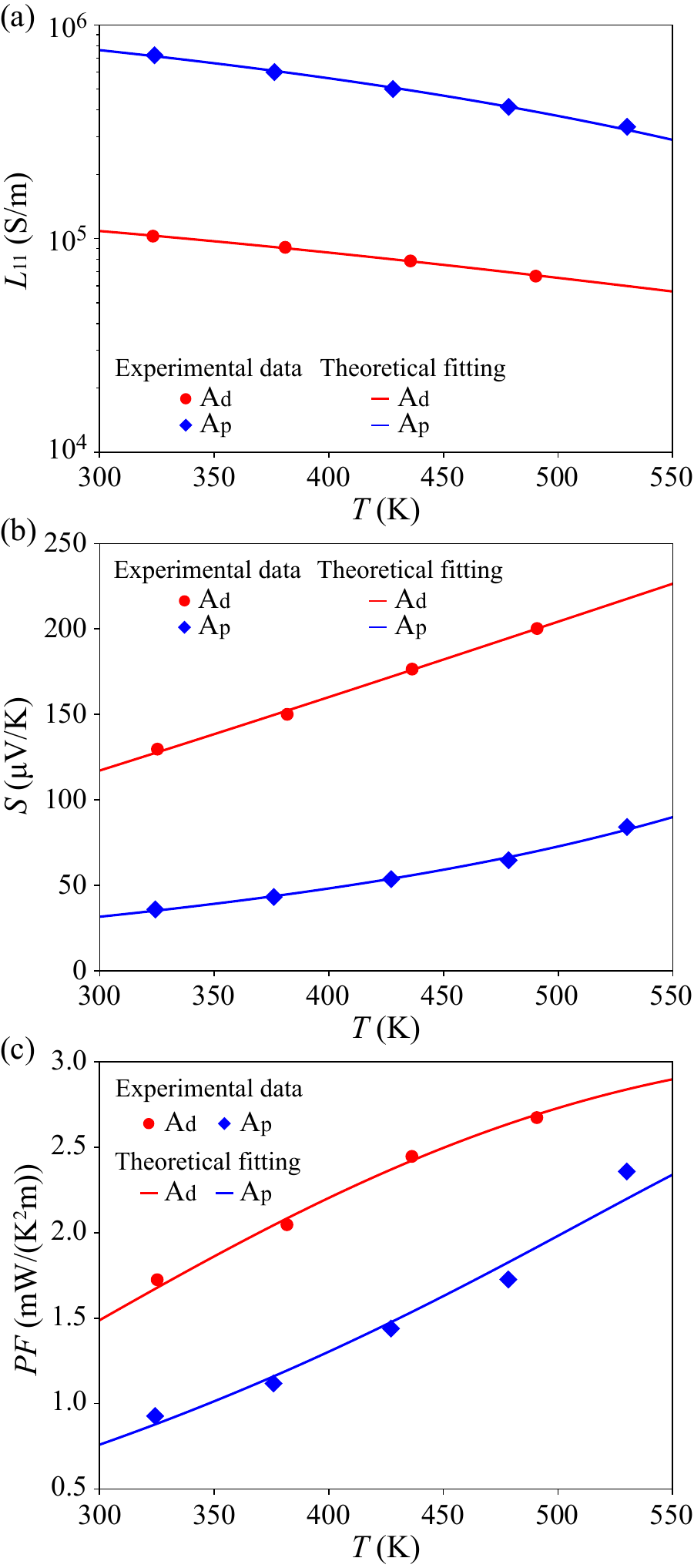}
    \end{center}
    \caption{(Color online) 
    $T$ dependence of (a) $L_{11}$, (b) $S$, and (c) $PF$ for pristine ($\rm A_p$ (blue)) and doped ($\rm A_d$ (red)) GeTe. The squares and circles correspond to experimental data~\cite{rf:gete_mori} for $\rm A_p$ and $\rm A_d$, respectively, and solid curves are the results of model fitting.
    }
    \label{fig:gete_l11}
\end{figure}

$L_{11}$, $S$, and $PF$ for both pristine GeTe and doped GeTe (Ge$_{0.87}$Y$_{0.02}$Sb$_{0.10}$Ag$_{0.01}$Te), hereafter expressed as $\rm A_p$ and $\rm A_d$, respectively, are shown in Fig.~\ref{fig:gete_l11}.
{Here, we focus on the experimental data corresponding to the rhombohedral phase below 550~K because the cubic phase in $\rm A_d$ is formed at 573~K~\cite{rf:gete_mori}.}
The squares and circles in Fig.~\ref{fig:gete_l11} are experimental data~\cite{rf:gete_mori}, and the lines are the results of model fitting with the parameters given in Table~\ref{tb:gete}, with $l_{\rm el}$ being the mean free path of electrons at $300$~K deduced by $l_{\rm el}=\frac{\hbar}{\Gamma}\sqrt{\frac{\mu}{2m^*}}$.
{The fitting parameters, $\Gamma$ and $\mu$, are determined to reproduce the $T$ dependence of both $L_{11}$ and $S$ in the experiment~\cite{rf:gete_mori}, with the mean relative error (MRE) being less than 1.7$\%$ for both $L_{11}$ and $S$ of $\rm A_p$ and less than 0.8$\%$ for those of $\rm A_d$.}

\begin{table}[h]
    \centering
    \caption{
      Fitting parameters for GeTe-based compounds for $L_{11}$, $S$ and $PF$, and $l_{\rm el}$ at $300$~K.}
    \label{tb:gete}
    \begin{center}
        \begin{tabular}{l|rr}
            \hline & GeTe ($\rm A_p$) & Ge$_{0.87}$Y$_{0.02}$Sb$_{0.10}$Ag$_{0.01}$Te ($\rm A_d$)\\ 
            \hline \hline
            $\mu$~(eV) & {$\varepsilon_F$+93.5~($k_BT$)$^2$} & {$\varepsilon_F$+41.4~($k_BT$)$^2$}\\ 
            $\varepsilon_F$~(eV) & ${-4.05\times 10^{-1}}$ & ${-9.74\times 10^{-2}}$\\ 
            $\Gamma_{\rm imp}$~(eV) & ${2.45\times10^{-2}}$ & ${2.60\times10^{-2}}$\\ 
            $\gamma_{\rm el-ph}$~(eV/K) & ${3.04\times10^{-5}}$ & ${1.91\times10^{-6}}$\\
            $l_{\rm el}$~(nm) & ${6.79}$ & ${3.88}$\\  
            \hline
        \end{tabular}
    \end{center}
\end{table}

The Fermi energy $\varepsilon_F$ for $\rm A_p$ and $\rm A_d$ shown in Table~\ref{tb:gete} is consistent with the experimental data~\cite{rf:gete_mori} in view of the fact that the number of holes decreases upon doping by electrons from Y and Sb. As evident in Figs.~\ref{fig:gete_l11}(a) and \ref{fig:gete_l11}(b), $L_{11}$ is smaller and $S$ is larger in $\rm A_d$ than the corresponding values in $\rm A_p$, reflecting the difference in the number of carriers. 
The results of parameter fitting, $\Gamma_{\rm imp}$ and $\gamma_{\rm el-ph}$, are reasonable. Regarding $\Gamma_{\rm imp}$, it is natural that impurity scattering increased in the doped case, $\rm A_d$, with more planar defects and finer domain structures, whereas a possible explanation of our new finding of weaker phonon scattering $\gamma_{\rm el-ph}$ in more disordered cases is given in Appendix~\ref{sec:appendix}. The decrease in $\gamma_{\rm el-ph}$ suppresses the increase in $\Gamma$ at high $T$, resulting in a higher $PF$ in $\rm A_d$ than in $\rm A_p$. 
The estimated value of $l_{\rm el}$ in Table~\ref{tb:gete} is smaller than the grain size, indicating that the $PF$, which is mainly due to electrons, is influenced by the electronic states within the grain. This is in sharp contrast to $\kappa_{\rm ph}$, as will be discussed in  Section~\ref{sbsec:gete_zt}.

\begin{figure}[t]
    \begin{center}
        \includegraphics[keepaspectratio=true,width=76mm]{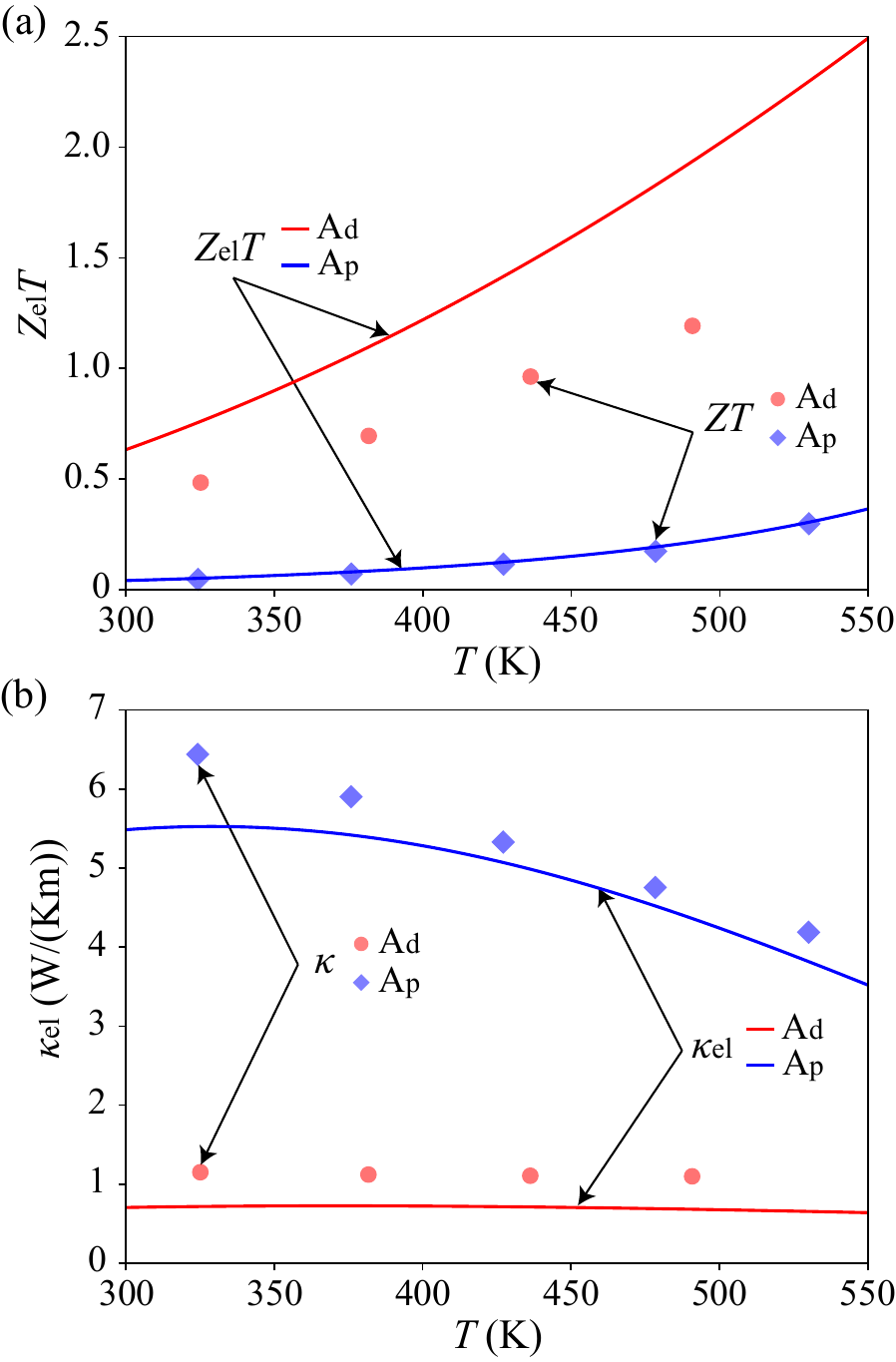}
    \end{center}
    \caption{(Color online)
    Solid curves show the $T$ dependence of (a) $Z_{\rm el}T$ and (b) $\kappa_{\rm el}$ for pristine ($\rm A_p$ (blue)) and doped ($\rm A_d$ (red)) GeTe, accounting only for the electronic contribution, as obtained from parameter fitting (Table~\ref{tb:gete}). The squares and circles correspond to (a) $ZT$ and (b) $\kappa$ observed experimentally~\cite{rf:gete_mori} for comparison.}
    \label{fig:gete_zt}
\end{figure}

Using parameters obtained rather uniquely through analysis of $L_{11}$ and $S$, it is possible to determine $L_{22}^{\rm el}$ and, subsequently, the electronic contribution to $ZT$ defined by $Z_{\rm el}T = \frac{PF}{\kappa_{\rm el}}T$, in terms of $\kappa_{\rm el} =\left(L_{22}^{\rm el}-\frac{L_{12}^2}{L_{11}}\right)\frac{1}{T}$. 
The $Z_{\rm el}T$ and $\kappa_{\rm el}$ are shown as solid curves in Figs.~\ref{fig:gete_zt}(a) and \ref{fig:gete_zt}(b), respectively, where experimental data~\cite{rf:gete_mori} for $ZT$ and $\kappa$ are included for comparison. As shown in Fig.~\ref{fig:gete_zt}(a), the difference between $Z_{\rm el}T$ and $ZT$ provides information about $\kappa_{\rm ph}/\kappa_{\rm el}$ because $Z_{\rm el}T-ZT= (\kappa_{\rm ph}/\kappa_{\rm el})ZT$. Using the $\kappa_{\rm el}$ curve in Fig.~\ref{fig:gete_zt}(b), it is now possible to identify $\kappa_{\rm ph}$.

\subsection{Phononic Contribution to Thermal Conductivity; $\kappa_{\rm ph}$~\label{sbsec:gete_zt}}
\begin{figure}[t]
    \begin{center}
        \includegraphics[keepaspectratio=true,width=76mm]{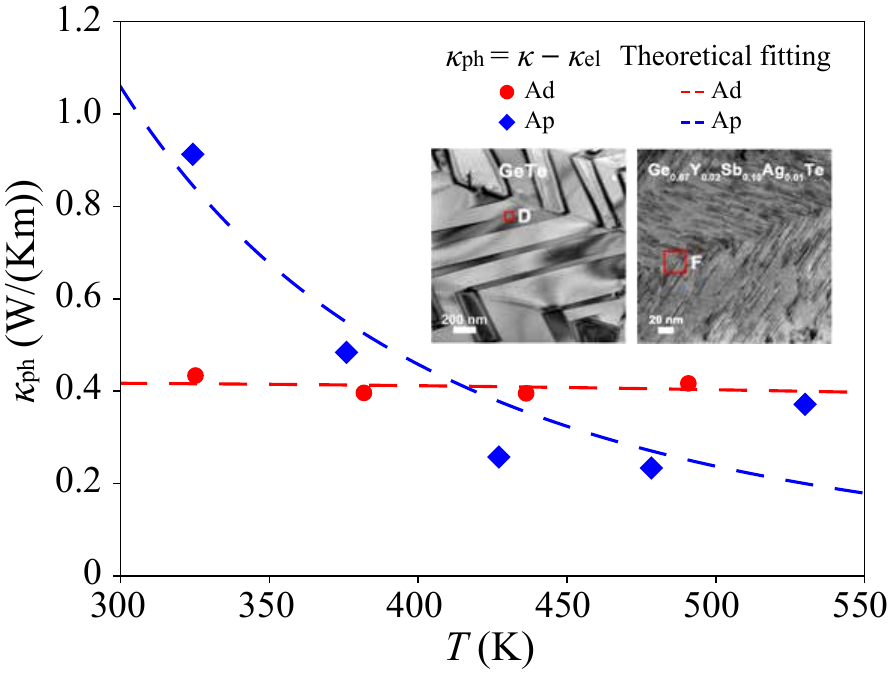}
    \end{center}
    \caption{(Color online)
    The squares and circles correspond to the $T$ dependences of $\kappa_{\rm ph} = \kappa - \kappa_{\rm el}$ for pristine ($\rm A_p$ (blue)) and doped ($\rm A_d$ (red)) GeTe, respectively; dashed curves are the results of a theoretical fitting based on the Holland formula in Eq.~(\ref{eq:holland}).
    {The insets on the left and right show transmission electron microscope (TEM) images of $\rm A_p$ and $\rm A_d$, respectively, reproduced from [6].
    Reprinted with permission from AAAS.
    }}
    \label{fig:gete_ph}
\end{figure}

In Fig.~\ref{fig:gete_ph}, the squares and circles represent $\kappa_{\rm ph}(=\kappa - \kappa_{\rm el})$, in ${\rm A_p}$ and ${\rm A_d}$, respectively. The new findings regarding the magnitude and $T$ dependences of $\kappa_{\rm ph}$, as disclosed in Fig.~\ref{fig:gete_ph}, indicate that $\kappa_{\rm ph}$ in ${\rm A_d}$ is less dependent on $T$ than $\kappa_{\rm ph}$ in ${\rm A_p}$.
We will analyze this result based on previous studies derived for insulators~\cite{rf:kappa_klemens,rf:kappa_callaway,rf:kappa_holland}.
(Notably, an extra contribution from electron--phonon scattering is present in conductors but is ignored here.)
For example, according to Holland, $\kappa_{\rm ph}$ is given by~\cite{rf:kappa_holland}
\begin{eqnarray}
    \kappa_{\rm ph}=\frac{k_B}{2\pi^2v}
    \left(\frac{k_BT}{\hbar}\right)^3
    \int_0^{\theta_D/T}
    \frac{x^4e^x(e^x-1)^{-2}}
    {v/L+a x^4T^4+b x^2T^5}dx,
    \label{eq:holland}
\end{eqnarray}
where scattering by boundaries ($v/L$) and impurities ($a$, due to differences in mass and spring constants) and scattering among phonons ($b$, due to three phonons) are taken into account. $\kappa_{\rm ph}$, expressed in Eq.~(\ref{eq:holland}), increases in proportion to $T^3$ at low $T$ and decreases at $T$ above the Debye temperature, $\theta_{\rm D}$, because of three-phonon interactions. Dashed curves in Fig.~\ref{fig:gete_ph} show the results of theoretical fitting of $\kappa_{\rm ph} = \kappa - \kappa_{\rm el}$ with $L$, $a$, and $b$ in the Holland formula (Eq.~(\ref{eq:holland})). The fitting parameters listed in Table~\ref{tb:gete_ph} were obtained under the assumptions that the sample size, $L$, is the average grain size in the experiment~\cite{rf:gete_mori}, the average phonon velocity, $v$, is 2100~m/s, and $\theta_{\rm D} = 199$~K, which are values reported for pristine GeTe~\cite{rf:gete_v1,rf:gete_v2}.
{
It is seen that in $\rm A_p$, 
the $T$ dependence of $\kappa_{\rm ph}$ is essentially determined by $b$, rather than $a$, although the value of $b$ is not highly reliable due to the MRE of 25.9$\%$.
}
On the other hand, in $\rm A_d$, in which inhomogeneity is greater because of an increase in the distribution of planar defects and smaller-sized grains, $v/L$ is larger and $a$ is more effective,
{with the lower MRE of 3.7$\%$}.
A microscopic understanding of these interesting results between $\kappa_{\rm ph}$ and shapes/sizes of grains is beyond the scope of the present paper. However, we note that the Boltzmann type phenomenological treatment $\kappa_{\rm ph}=\frac{1}{3}Cvl_{\rm ph}$, where $C$ (=1.5~J/(cm$^3$K) at 300~K~\cite{rf:gete_c}) is the specific heat and $l_{\rm ph}$ is the mean free path of phonons, results in $l_{\rm ph}$ = 1.01~nm and 0.40~nm for $\rm A_p$ and $\rm A_d$, respectively, at 300~K. Thus, $l_{\rm ph}$ is considerably smaller than the grain size and also smaller than $l_{\rm el}$ shown in Table~\ref{tb:gete}, which is not physically acceptable. This unphysical result is attributable to the fact that the effects of $a$ and $b$ in Eq.~(\ref{eq:holland}) are not properly taken into account because of their dependences on $x$, which are variables to be integrated.

\begin{table}[h]
    \centering
    \caption{
      Fitting parameters for GeTe-based compounds for $\kappa_{\rm ph}$}
    \label{tb:gete_ph}
    \begin{tabular}{l|rr}
    \hline& GeTe ($\rm A_p$) & Ge$_{0.87}$Y$_{0.02}$Sb$_{0.10}$Ag$_{0.01}$Te ($\rm A_d$)\\ 
    \hline \hline
    $L$~($\mu$m) & {$3.28\times10^{-1}$} & {$3.17\times10^{-1}$}\\
    $a$~(s$^{-1}$K$^{-4}$) & {$\sim0$} & {1.$19\times10^{5}$}\\ 
    $b$~(s$^{-1}$K$^{-5}$) & {$4.86$} & {$4.02\times10^{-2}$}\\ 
    \hline
  \end{tabular}
\end{table}

\section{Case of Mg$_3$Sb$_2$~\label{sec:mgsb}}

\subsection{Electronic States and Model of Spectral Conductivity}
\begin{figure}[t]
    \begin{center}
    \includegraphics[keepaspectratio=true,width=76mm]{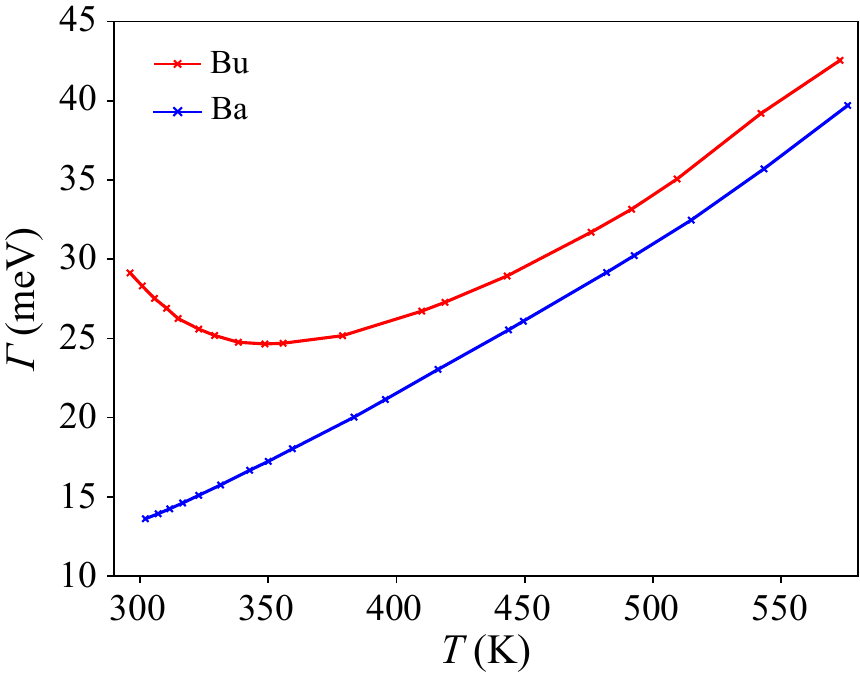}
    \end{center}
    \caption{(Color online)
    $T$ dependence of fitting-parameter $\Gamma$ for unannealed ($\rm B_u$ (red)) and annealed ($\rm B_a$ (blue)) Mg$_{3+\delta}$Sb$_{1.49}$Bi$_{0.5}$Te$_{0.01}$ samples~\cite{rf:mgsb_imasato}.}
    \label{fig:gamma_mgsb}
\end{figure}

Mg$_3$Sb$_2$ has a trigonal crystal structure, and its conduction-band minimum is in the L$^*$-M$^*$ line with $g_{\rm v}=6$, which is dominated by the electronic state of the Mg atoms~\cite{rf:mgsb_band,rf:mgsb_orbital}.
In the case of Mg$_3$Sb$_2$-based compounds with Sb site substitution, Mg$_{3+\delta}$Sb$_{1.49}$Bi$_{0.5}$Te$_{0.01}$~\cite{rf:mgsb_imasato}, the electrons are doped with $m^* = 0.235m_0$ (the anisotropy of $m^*$ is neglected)~\cite{rf:mgsb_mass,rf:mgsb_mass2}.
In contrast to the case of GeTe, where doping effects are analyzed, we here focus on the annealing-induced change in the thermoelectric performance of a specific sample of Mg$_{3+\delta}$Sb$_{1.49}$Bi$_{0.5}$Te$_{0.01}$ because it shows clear and characteristic variations associated with grain size. The grain sizes in unannealed and annealed samples were $\sim10$~$\mu$m and $>30~\mu$m, respectively~\cite{rf:mgsb_imasato}.
Unannealed and annealed samples of Mg$_{3+\delta}$Sb$_{1.49}$Bi$_{0.5}$Te$_{0.01}$ are hereafter represented as $\rm B_u$ and $\rm B_a$, respectively.
$L_{11}$ in $\rm B_a$ shows a monotonic decrease with increasing $T$, indicating that phonon scattering is dominant,
whereas $L_{11}$ in $\rm B_u$ exhibits a non-monotonic $T$ dependence, suggesting that other scattering mechanisms (e.g., scattering by grain boundaries and ionized impurities) may be activated.
Hence, $\Gamma$ is taken as a fitting parameter to capture the feature of $L_{11}$ by a similar scheme to $\alpha(\varepsilon)$ in Eq.~(\ref{eq:alpha}) in the case of GeTe. Fig.~\ref{fig:gamma_mgsb} and Table~\ref{tb:mgsb} show the fitting parameters $\Gamma$ and $\mu$ found to reproduce the $T$ dependence of both $L_{11}$ and $S$ in the experiments~\cite{rf:mgsb_imasato}, {yielding the MRE~$<1.1\%$ for both $L_{11}$ and $S$ of $\rm B_u$ and $\rm B_a$.}
The results show that $\Gamma$ is proportional to $T$ in the high-$T$ region, in accordance with the expected dominance of electron--phonon scattering for both $\rm B_u$ and $\rm B_a$, while $\Gamma$ in $\rm B_u$ deviates from $T$ in the low-$T$ region because of the occurrence of other scattering mechanisms.

\begin{table}[h]
    \centering
    \caption{
      Fitting parameters for Mg$_3$Sb$_2$-based compounds for $L_{11}$, $S$, $PF$, and $l_{\rm el}$ at $300$~K}
    \label{tb:mgsb}
    \begin{tabular}{l|rr}
    \hline& Annealed ($\rm B_a$) & Unannealed ($\rm B_u$)\\
    \hline \hline
    $\mu$~(eV) & {$\varepsilon_F$-24.4~($k_BT$)$^2$} & {$\varepsilon_F$-28.0~($k_BT$)$^2$} \\ 
    $\varepsilon_F$~(eV) & {$3.95\times 10^{-2}$} & {$4.23\times 10^{-2}$} \\ 
    $l_{\rm el}$~(nm) & {4.56} & {2.17} \\  \hline
    \end{tabular}
\end{table}

\subsection{Parameter-Fitting for $L_{11}$, $S$, $PF$, and Electronic Contributions to $ZT$}
\begin{figure}[t]
    \begin{center}
    \includegraphics[keepaspectratio=true,width=76mm]{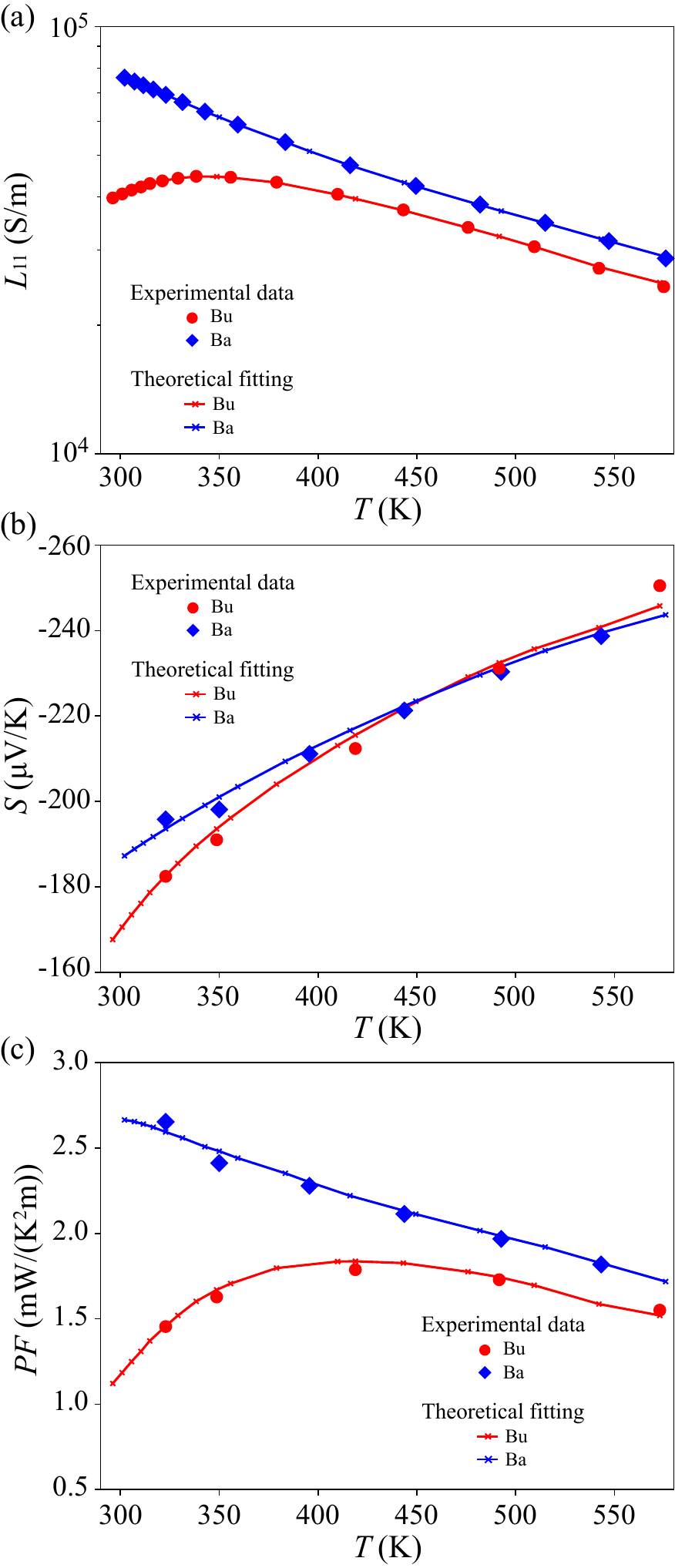}
    \end{center}
    \caption{(Color online) 
    $T$ dependences of (a) $L_{11}$, (b) $S$, and (c) $PF$ for unannealed ($\rm B_u$ (red)) and annealed ($\rm B_a$ (blue)) Mg$_{3+\delta}$Sb$_{1.49}$Bi$_{0.5}$Te$_{0.01}$. The circles and squares correspond to experimental data~\cite{rf:mgsb_imasato} for $\rm B_u$ and $\rm B_a$, respectively; solid curves are the model fitting results.}
    \label{fig:mgsb_l11}
\end{figure}

\begin{figure}[t]
    \begin{center}
    \includegraphics[keepaspectratio=true,width=76mm]{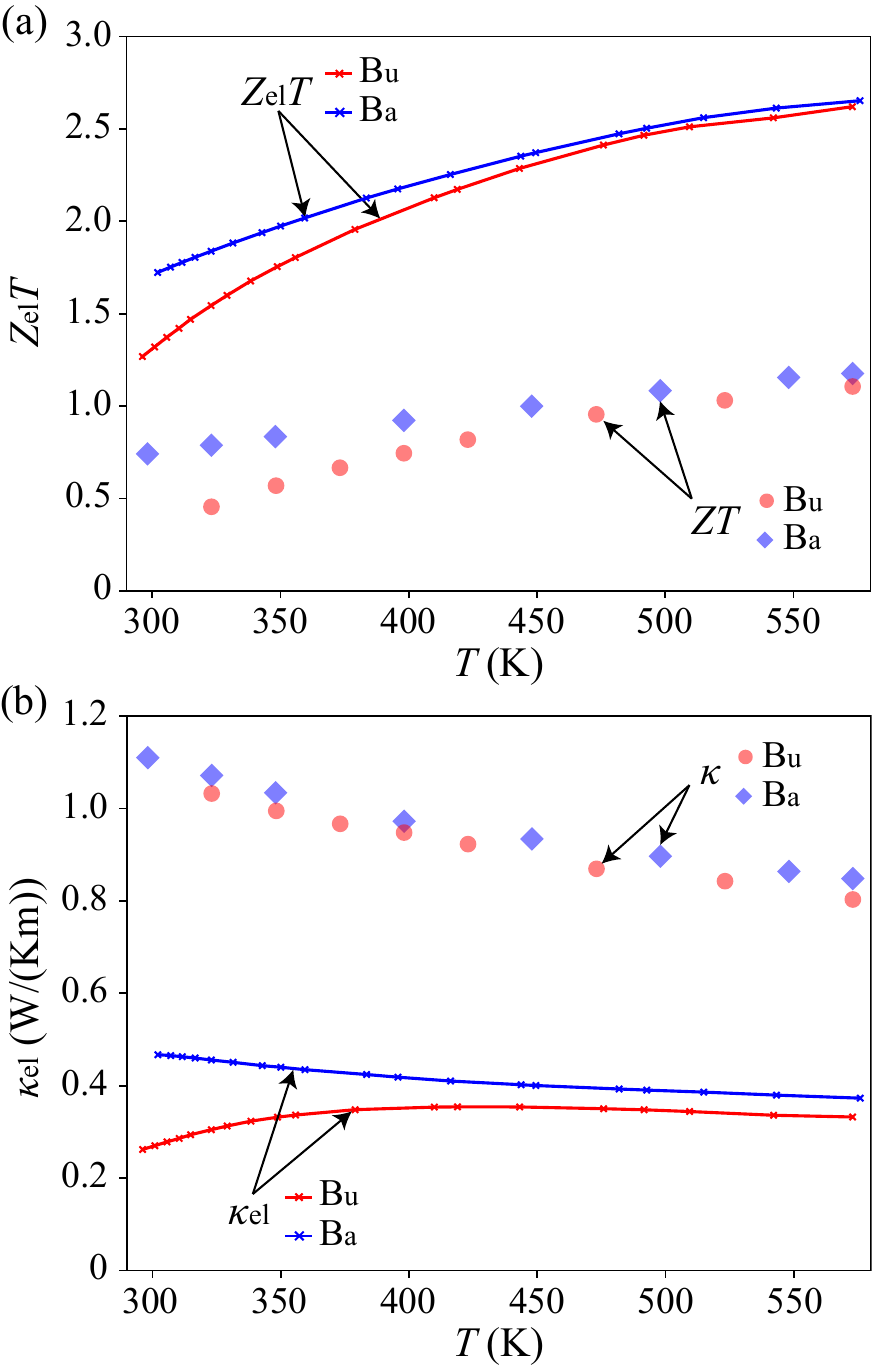}
    \end{center}
    \caption{(Color online) 
    Solid curves show the $T$ dependence of (a) $Z_{\rm el}T$ and (b) $\kappa_{\rm el}$, accounting only for the {electronic} contribution, as obtained from parameter fitting in Fig.~\ref{fig:gamma_mgsb} and Table~\ref{tb:mgsb}. The circles and squares correspond to (a) $ZT$ and (b) $\kappa$ observed experimentally~\cite{rf:mgsb_imasato} for comparison.}
   \label{fig:mgsb_zt}
\end{figure}

Plots of $L_{11}$, $S$, and $PF$ as functions of $T$ are shown for $\rm B_u$ and $\rm B_a$ in Fig.~\ref{fig:mgsb_l11}, where squares and circles are experimental data~\cite{rf:mgsb_imasato} and solid curves are the results of model fitting with the parameters given in Fig.~\ref{fig:gamma_mgsb} and Table~\ref{tb:mgsb}. Reflecting the $\Gamma$ behavior, the $T$ dependences of both $L_{11}$ and $S$ are larger in $\rm B_a$, resulting in a larger $PF$, especially at low $T$. 
As in the case of GeTe, we extracted $\kappa_{\rm el}$ from the results of the parameter fittings of $L_{11}$, $S$, and $PF$; comparisons of the $Z_{\rm el}T$ and $\kappa_{\rm el}$ with the experimentally observed $ZT$ and $\kappa$ are shown in Figs.~\ref{fig:mgsb_zt}~(a) and \ref{fig:mgsb_zt}~(b), respectively. The results show that, at low $T$, both $ZT$ and $Z_{\rm el}T$ increase as a result of annealing.

\subsection{Phononic Contribution to Thermal Conductivity, $\kappa_{\rm ph}$}
In the case of pristine Mg$_3$Sb$_2$, Kanno et al. indicated that $\kappa_{\rm ph}$ is proportional to $T^2$ at low $T$ and decreases moderately with increasing $T$ ($\kappa_{\rm ph}\sim T^{-0.5}$) at high $T$~\cite{rf:mgsb_kanno}.

\begin{figure}[t]
    \begin{center}
    \includegraphics[keepaspectratio=true,width=76mm]{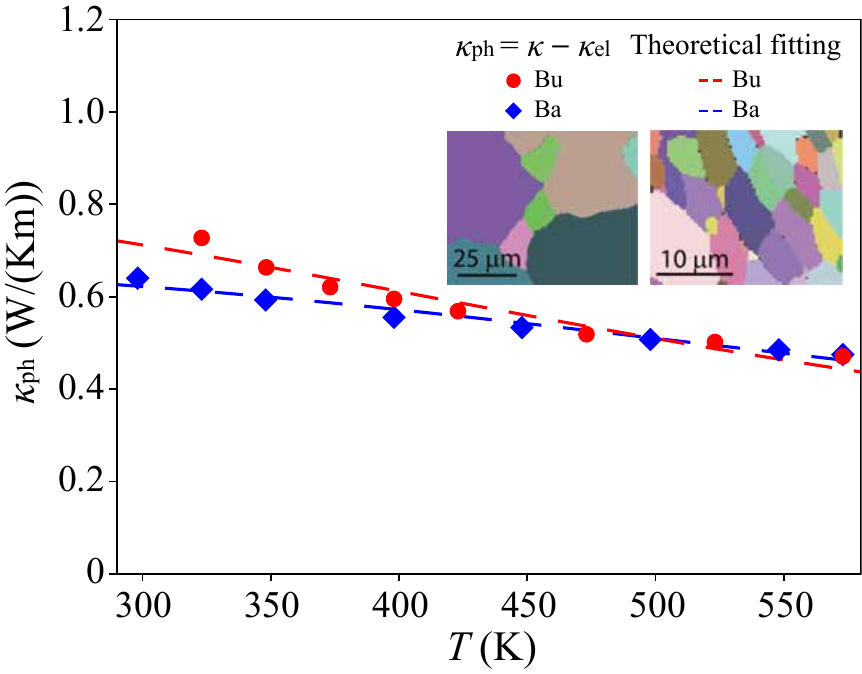}
    \end{center}
    \caption{(Color online)
    The circles and squares correspond to the $T$ dependence of $\kappa_{\rm ph}=\kappa-\kappa_{\rm el}$ for unannealed ($\rm B_u$ (red)) and annealed ($\rm B_a$ (blue)) Mg$_{3+\delta}$Sb$_{1.49}$Bi$_{0.5}$Te$_{0.01}$, respectively; dashed curves are the results of theoretical fitting based on the Holland formula in Eq.~(\ref{eq:holland}).
    {The insets on the left and right show electron backscatter diffraction (EBSD) maps for $\rm B_a$ and $\rm B_u$, respectively.
    Used with permission of John Wiley \& Sons-Books, from [7];
    permission conveyed through Copyright Clearance Center, Inc.}
    }
    \label{fig:mgsb_ph}
\end{figure}

In the present case, $\kappa_{\rm ph}$ is shown in Fig.~\ref{fig:mgsb_ph}, with the results of parameter fitting in $\kappa_{\rm ph}$ based on Eq.~(\ref{eq:holland}) given in Table~\ref{tb:mgsb_ph} under the assumption that $L$ is the average grain size in the experiment~\cite{rf:mgsb_imasato}, $\theta_{\rm D} = 209$~K, and $v = 2076$~m/s~\cite{rf:mgsb_v2}, {with the MRE of 1.8$\%$ and 3.4$\%$ for $\kappa_{\rm ph}$ of $\rm B_a$ and $\rm B_u$, respectively}.
The results indicate that the effect of increasing grain size by annealing on $\kappa_{\rm ph}$ is globally minor except for the decrease in $\kappa_{\rm ph}$ at low $T$.

\begin{table}[h]
    \centering
    \caption{Fitting parameters for Mg$_3$Sb$_2$-based compounds for $\kappa_{\rm ph}$}
    \label{tb:mgsb_ph}
    \begin{tabular}{l|rr}
    \hline& Annealed ($\rm B_a$) & Unannealed ($\rm B_u$)\\
    \hline \hline 
    $L$~($\mu$m) & 30 & 10\\ 
    $a$~(s$^{-1}$K$^{-4}$) & {3.23 $\times 10^5$} & {1.67$\times$10$^5$} \\ 
    $b$~(s$^{-1}$K$^{-5}$) & {5.31 $\times 10^{-2}$} & {1.46$\times 10^{-1}$} \\ 
    \hline
\end{tabular}
\end{table}

\section{Comparison between GeTe and Mg$_3$Sb$_2$ \label{sec:comparison}}
Both GeTe and Mg$_3$Sb$_2$ are high-$ZT$ materials because of their low $\kappa$. A comparison reveals that their $\kappa_{\rm ph}$ values do not substantially differ, although the average grain size in GeTe is almost two orders of magnitude smaller than that in Mg$_3$Sb$_2$ according to experiments~\cite{rf:gete_mori,rf:mgsb_imasato}.
By contrast, the $T$ dependence of $\kappa_{\rm ph}$ differs substantially between GeTe and Mg$_3$Sb$_2$.
In the case of GeTe, the $T$ dependence of $\kappa_{\rm ph}$ is strong for pristine GeTe ($\rm A_p$) but weak for doped GeTe ($\rm A_d$) with a finer domain structures.
However, $\kappa_{\rm ph}$ is almost independent of grain size in the case of Mg$_3$Sb$_2$.
This difference is attributable to differences in the grain shapes, disrupted herringbone structure type irregular one in GeTe {shown in the inset of Fig.~\ref{fig:gete_ph}} vs. rather uniform amorphous like one in Mg$_3$Sb$_2$ {in the inset of Fig.~\ref{fig:mgsb_ph}.}

This result indicates that not only the sizes but also the shapes of grains should be optimized to minimize $\kappa_{\rm ph}$, leading to a higher $ZT$.

\section{Summary \label{sec:summary}}
The Sommerfeld--Bethe formula interrelates $L_{11}$ (electrical conductivity), $L_{12}$ (thermoelectrical conductivity), and $L_{22}^{\rm el}$ (contribution of electrons to thermal conductivity) via a common physical quantity: spectral conductivity, $\alpha$.
In the present paper, we proposed that a detailed analysis of experimental data for both the conductivity, $L_{11}$, and Seebeck coefficient, $S=\frac{1}{T}\frac{L_{12}}{L_{11}}$, and then power factor, $PF=\frac{1}{T^2}\frac{L_{12}^2}{L_{11}}$, can result in unique dependences on energy and temperature of $\alpha$, which in turn can lead to the identification of the contribution of electrons to the thermal conductivity, $\kappa_{\rm el}=\left(L_{22}^{\rm el}-\frac{L_{12}L_{21}}{L_{11}}\right)\frac{1}{T}$.
By comparing {$\kappa_{\rm el}$ with} experimental data for thermal conductivity, $\kappa$, governing $ZT=\frac{PF}{\kappa}T$, we infer that the {phononic} thermal conductivity, $\kappa_{\rm ph}=\kappa-\kappa_{\rm el}$, which has been analyzed on the basis of the formula by Holland. 
{This analysis indicates the dominant roles of scattering by boundaries, impurities, and three phonons.}
To the best of our knowledge, this work is the first to identify various sources of $T$ dependences of $\kappa_{\rm ph}$.

GeTe and Mg$_3$Sb$_2$, which are known to exhibit high $ZT$, are inhomogeneous and {have different morphologies}: GeTe exhibits a disrupted herringbone structure, and Mg$_3$Sb$_2$ shows a rather uniform amorphous structure. 
In GeTe, the $T$ dependence of $\kappa_{\rm ph}$ is sensitive to grain size and defects; by contrast, in Mg$_3$Sb$_2$, $\kappa_{\rm ph}$ is insensitive to grain size. These results indicate that not only the size but also the shapes of grains affect $\kappa_{\rm ph}$, which deserves more systematic studies.

Based on the present results, detailed observations of sample structures and analyses of grain patterns (e.g., by TEM and image analysis), along with spectroscopic information (e.g., X-ray absorption or emission and photoemission spectroscopy) to identify the density of states (DOS), which are useful for constructing a model of $\alpha$, will be effective for further exploring higher-$ZT$ materials.

It is of interest to extend the present analysis to Mg$_3$(Sb,Bi)$_2$ based materials, which are expected to have practical importance as useful thermoelectrics~\cite{rf:mgsbbi_mori}.

\begin{acknowledgment}
We thank Takao Mori (NIMS) for fruitful discussions on experiments with GeTe.
\end{acknowledgment}

\onecolumn

\appendix
\section{Effects of Disorder on Phonon Scattering at High Temperature \label{sec:appendix}}
\begin{figure}[t]
  \begin{center}
  \includegraphics[keepaspectratio=true,width=40mm]{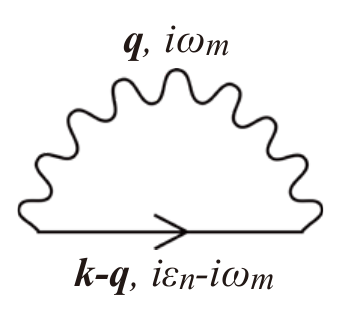}
  \end{center}
\caption{Diagram of self-energy for electron scattering by phonons.
The solid and wavy lines represent the electron and phonon, respectively.
}
\label{fig:el_ph}
\end{figure}

The effects of phonon scattering on electrons are represented by the self-energy corrections in Fig.~\ref{fig:el_ph}, where solid and wavy lines are electron and phonon Green's functions, respectively.
In clean systems, this process is estimated as follows~\cite{rf:bloch}:
\begin{align}
  \Sigma^{\rm R}_{\rm el-ph}({\bm k},\varepsilon)
  &=-\frac{k_BT}{V}\sum_{m, {\bm q}}g_{\bm q}^2\mathcal{D}({\bm q},i\omega_m)\mathcal{G}({\bm k}-{\bm q},i\varepsilon_n-i\omega_m)|_{i\varepsilon_n\rightarrow\varepsilon+i\delta}\notag\\
  &=-\frac{1}{V}\sum_{\bm q}g_{\bm q}^2
  \left[
    \frac{f(\xi_{{\bm k}-{\bm q}})-1-n(\hbar \omega_{\bm q})}{\varepsilon-\xi_{{\bm k}-{\bm q}}-\hbar\omega_{\bm q}+i\delta}
    -\frac{f(\xi_{{\bm k}-{\bm q}})+n(\hbar \omega_{\bm q})}{\varepsilon-\xi_{{\bm k}-{\bm q}}+\hbar\omega_{\bm q}+i\delta}
 \right]
\end{align}
where $\xi_{{\bm k}-{\bm q}}=\varepsilon_{{\bm k}-{\bm q}}-\mu$, $n(\hbar\omega_{\bm q})=1/(\exp(\beta\hbar\omega_{\bm q})-1)$ is the Bose distribution function, and $g_{\bm q}$ is the electron--phonon coupling constant $(g_{\bm q}^2=g q)$.
$\mathcal{D}({\bm q},i\omega_m)$ is the phonon Green's function, 
\begin{align}
  \mathcal{D}({\bm q},i\omega_m)=\frac{2\hbar\omega_{\bm q}}{(i\omega_m)^2-(\hbar\omega_{\bm q})^2},
\end{align}
and $\mathcal{G}({\bm k}-{\bm q},i\varepsilon_n-i\omega_m)$ is the free-electron Green's function,
\begin{align}
  \mathcal{G}({\bm k}-{\bm q},i\varepsilon_n-i\omega_m)=\frac{1}{i\varepsilon_n-i\omega_m-\xi_{{\bm k}-{\bm q}}}.
\end{align}
The real part of $\Sigma^{\rm R}_{\rm el-ph}({\bm k},\varepsilon)$ leads to a well-known result at $T=0$, where $n(\hbar\omega_{\bm q})=0$, determining the mass enhancement of the electronic specific heat at low $T$.
At high $T$, $k_BT\gg\hbar\omega_D$ ($\omega_D$: Debye frequency), where $n(\hbar\omega_{\bm q})\sim k_BT/\hbar\omega_{\bm q}$, the imaginary part of $\Sigma^{\rm R}_{\rm el-ph}({\bm k},\varepsilon)$ at $\varepsilon=0$ leads to $-\frac{g\omega_D^2}{4\pi\hbar^2 v^3v_F}k_BT$, where $v_F$ is the Fermi velocity. 

However, in the present disordered metals, $\Sigma^{\rm R}_{\rm el-ph}({\bm k},\varepsilon)$ is rewritten using the electron Green's function with disorder, $\gamma_{\rm el}$, together with the phonon propagator with damping, $\gamma_{\rm ph}$, as follows:
\begin{align}
  \Sigma^{\rm R}_{\rm el-ph}({\bm k},\varepsilon)
  =-\frac{k_BT}{V}\sum_{m, {\bm q}}g_{\bm q}^2
  \frac{1}{i\varepsilon_n-i\omega_m-\xi_{{\bm k}-{\bm q}}+i\gamma_{\rm el}{\rm sign}(\varepsilon_n-\omega_m)}
  \left.\left(
    \frac{1}{i\omega_m-\hbar\omega_{\bm q} + i\gamma_{\rm ph}{\rm sign}(\omega_m)} 
    - \frac{1}{i\omega_m+\hbar\omega_{\bm q}+i\gamma_{\rm ph}{\rm sign}(\omega_m)}
  \right)\right|_{i\varepsilon_n\rightarrow\varepsilon+i\delta}.
\end{align}
At high $T$, the imaginary part of $\Sigma^{\rm R}_{\rm el-ph}({\bm k},\varepsilon)$ is given as
\begin{align}
  {\rm Im}\Sigma^{\rm R}_{\rm el-ph}({\bm k},\varepsilon)
  \sim-\frac{2k_BT}{V}
  \sum_{{\bm q}}g_{\bm q}^2\hbar\omega_{\bm q}
  \frac{
    (\gamma_{\rm el}+2\gamma_{\rm ph})
    \left\{
      (\hbar\omega_{\bm q})^2 + (\gamma_{\rm el}+\gamma_{\rm ph})^2
    \right\}
    +\gamma_{\rm el}(\varepsilon-\xi_{{\bm k}-{\bm q}})^2
  }{
    \left\{
      (\hbar\omega_{\bm q})^2+\gamma_{\rm ph}^2
    \right\}
    \left\{
      (\varepsilon - \xi_{{\bm k}-{\bm q}} - \hbar\omega_{\bm q})^2
      +(\gamma_{\rm el}+\gamma_{\rm ph})^2
    \right\}
    \left\{
      (\varepsilon-\xi_{{\bm k}-{\bm q}} + \hbar\omega_{\bm q})^2
      +(\gamma_{\rm el}+\gamma_{\rm ph})^2
    \right\}
  },
\end{align}
and in the disorder case ($\gamma_{\rm el}\gg\gamma_{\rm ph},\hbar\omega_D$), ${\rm Im}\Sigma^{\rm R}_{\rm el-ph}({\bm k},\varepsilon)$ is written as
\begin{align}
  {\rm Im}\Sigma^{\rm R}_{\rm el-ph}({\bm k},\varepsilon)
  &\sim -\frac{2k_BT}{V}\sum_{{\bm q}}g_{\bm q}^2\frac{\hbar\omega_{\bm q}}{\gamma_{\rm el}(\hbar^2\omega_{\bm q}^2+\gamma_{\rm ph}^2)}\notag\\
  &=-\frac{g(\hbar\omega_{\rm D})^3}{\pi^2(\hbar v)^4\gamma_{\rm el}}\left\{
    \frac{1}{3}
    -\frac{\gamma_{\rm ph}^2}{(\hbar\omega_{\rm D})^2}
    +\frac{\gamma_{\rm ph}^3}{(\hbar\omega_{\rm D})^3}\arctan\left(\frac{\hbar\omega_{\rm D}}{\gamma_{\rm ph}}\right)
  \right\}k_BT,
\end{align}
where $\hbar\omega_{\bm q}=\hbar vq$.
Thus, $\gamma_{\rm el}$ suppresses phonon scattering; in addition, $|{\rm Im}\Sigma^{\rm R}_{\rm el-ph}|$ in dirty metals is smaller than that in clean metals at high $T$ under the conditions that $\frac{\hbar\omega_{\rm D}}{\gamma_{\rm el}}<\frac{3\pi}{4}\frac{v}{v_F}$ and $\hbar\omega_{\rm D}\gg\gamma_{\rm ph}$, resulting in a smaller contribution of electron--phonon scattering in metals with stronger disorder.

\twocolumn

\end{document}